\documentclass[journal]{IEEEtran}

\usepackage[numbers,sort&compress,square]{natbib}
\usepackage{comment}
\usepackage{hyperref}
\usepackage{amsmath,amssymb,amsfonts}
\usepackage{upgreek}
\usepackage{graphicx}
\usepackage{textcomp}
\usepackage{xcolor}
\usepackage{tabu,multirow}
\usepackage{mathtools}
\usepackage{tabularx}
\usepackage{ragged2e} 
\usepackage{stfloats}

\def\BibTeX{{\rm B\kern-.05em{\sc i\kern-.025em b}\kern-.08em
    T\kern-.1667em\lower.7ex\hbox{E}\kern-.125emX}}

\newcommand{\quotes}[1]{``#1''}

\newcolumntype{Y}{>{\centering\arraybackslash}X}

\begin{document}

\title{Statistical Analysis of Geometric Algorithms in Vehicular Visible Light Positioning\\}

\author{Burak~Soner,~\IEEEmembership{Member,~IEEE,} and~Sinem~Coleri,~\IEEEmembership{Senior Member,~IEEE,}
\thanks{Burak Soner and Sinem Coleri are with the Department of Electrical and Electronics Engineering, Koc University, 34450 Istanbul, Turkey (e-mail: \{bsoner16, scoleri\}@ku.edu.tr). The authors acknowledge the support of CHIST-ERA grant CHISTERA-18-SDCDN-001, the Scientific and Technological Council of Turkey 119E350 and Ford Otosan.}}

\maketitle

\begin{abstract}

Vehicular visible light positioning (VLP) methods find relative locations of vehicles by estimating the positions of intensity-modulated head/tail lights of one vehicle (target) with respect to another (ego). Estimation is done in two steps: 1) relative bearing or range of the transmitter-receiver link is measured over the received signal on the ego side, and 2) target position is estimated based on those measurements using a geometric algorithm that expresses position coordinates in terms of the bearing-range parameters. The primary source of statistical error for these non-linear algorithms is the channel noise on the received signals that contaminates parameter measurements with varying levels of sensitivity. In this paper, we present two such geometric vehicular VLP algorithms that were previously unexplored, compare their performance with state-of-the-art algorithms over simulations, and analyze theoretical performance of all algorithms against statistical channel noise by deriving the respective Cramer-Rao lower bounds. The two newly explored algorithms do not outperform existing state-of-the-art, but we present them alongside the statistical analyses for the sake of completeness and to motivate further research in vehicular VLP. Our main finding is that direct bearing-based algorithms provide higher accuracy against noise for estimating lateral position coordinates, and range-based algorithms provide higher accuracy in the longitudinal axis due to the non-linearity of the respective geometric algorithms. 

\end{abstract}


\section{Introduction}

Vehicular visible light positioning (VLP) methods utilize visible light communication (VLC) signals from modulated vehicle head/tail LED lights for positioning purposes \cite{roadmap_vlp}. They are envisioned as a complementary solution to the readily-available sensor-based autonomous vehicle localization stack, primarily for collision avoidance and platooning applications \cite{depontemuller}. The high directionality of the vehicle light sources and the short distances between two vehicles during the intended application scenarios minimizes chances of strong multi-path components in the VLC channel \cite{vlc_channel_wnl} and enables cm-level positioning accuracy with VLP \cite{vvlc_survey_memedi}. Moreover, since received VLC signals are used (1D signals over time) rather than high-dimensional input data like images or 3D point clouds, algorithm complexity is extremely low in VLP compared to sensor-based systems (e.g., camera-based object detectors), enabling kHz-level positioning rates with moderate processing hardware \cite{soner_tvt}. These promises make VLP a suitable complementary addition to the autonomous vehicle localization stack alongside sensor-based technologies.

Vehicular VLP methods are two-step estimators: 1) relative bearing (angle) or range (distance) between a transmitter (TX) and one or more receivers (RX) are measured, and 2) an algorithm combines these measurements to estimate relative TX position. High vehicle mobility causes a bias in the estimation due to finite rate and latency \cite{trj_pred, prob_trj_pred}, but this bias shrinks significantly when estimation rates are higher than 50 Hz and estimation error gets dominated by statistical error due to lower signal-to-noise ratio (SNR) on the received VLC signal \cite{soner_tvt}. Bearing-range measurement techniques are typically tasked with reducing the effect of this VLC channel noise on their outputs, and positioning algorithms simply combine these measurements by applying appropriate geometric relations \cite{becha_ranging}. However, since the geometric relations are highly non-linear, the parameter measurement noise triggers different levels of position estimation error for each algorithm. The literature on such \quotes{geometric} VLP algorithms is still developing. Specifically, statistical analysis is necessary to identify which algorithm provides the best performance under possible conditions. Furthermore, there are still feasible geometric algorithms that are unexplored in the literature. 

In this paper, we formulate two previously unexplored geometric algorithms that can be used for vehicular VLP, namely, differential bearing-based positioning and single-receiver range-based positioning, and we analyze and compare the theoretical performance of the newly proposed algorithms with the current state-of-the-art algorithms. The rest of the paper is organized as follows: Vehicular VLP is described and current state-of-the-art algorithms are briefly reviewed in Section II, the two new geometric algorithms are presented in Section III, statistical analysis and related simulations are presented in Section IV, and our conclusions are summarized in Section V.

\section{Vehicular VLP Algorithms}

The current state-of-the-art in vehicular VLP uses geometric positioning algorithms, i.e., algorithms that simply invert geometric relations that tie TX-RX bearing and/or range parameters to lateral and longitudinal position coordinates. These algorithms provide high accuracy owing to the strong correlation between the received signal amplitude / phase and the TX-RX bearing / range that emerges from the line-of-sight (LoS) dominant VLC channel at distances relevant to collision avoidance and platooning \cite{vlc_channel_wnl_ml, vlc_channel_wnl}. This property, coupled with the fact that the input space dimensionality is low (max. three \cite{turan_3tx} TX-RX units are used and sample buffers are kept small to maximize rate) makes simple geometric algorithms a more suitable choice compared to higher-complexity methods like least squares \cite{heidi_singleStep} or machine-learning \cite{ann_vlp_roadside} which utilize inference over many RX buffers. 

Algorithms can be categorized with respect to which parameter they utilize for position estimation: bearing, range, or both. Current state-of-the-art methods utilize direct measurements for only one of those parameters from one TX to two RX units as opposed to differential measurements (e.g., difference in range values from RX1 and RX2 to TX1) or a combination of them. The differential versions of range and bearing parameters are respectively defined as 

\begin{figure}[t]
	\centering
	\includegraphics[width=0.45\textwidth]{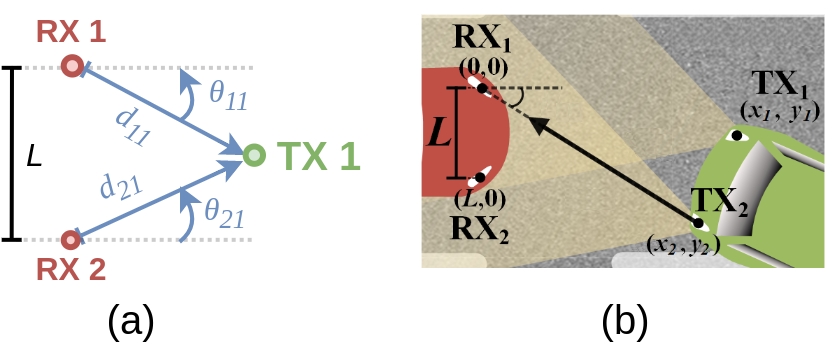}
	\vspace{0mm}
	\caption{(a) Abstract drawing of $\theta_{ij}, d_{ij} ~~i,j\in\{1,2\}$ which represent bearing and range values from RX $i$ to TX $j$ respectively, and (b) showing the same setup, with the link for TX 2 annotated on a vehicle. These quantities are estimated by the respective parameter measurement techniques to obtain $\widehat{\theta_{ij}}, \widehat{d_{ij}}$, which are used for positioning purposes.}
	\vspace{-2mm}
	\label{classicals}
\end{figure}

\vspace{-1mm}
\begin{equation}
	\label{diffs}
	\Delta d_{il/j} \!=\! d_{ij} \!- \! d_{lj}~~~~ \text{and}~~~~ \Delta \theta_{il/j}\! =\! \theta_{ij}\! -\! \theta_{lj}~~,
\end{equation}
\vspace{-2mm}

\noindent where $d_{ij}$ denotes range and $\theta_{ij},~~i,j \in \{1,2,3,4\}$ denotes bearing from RX $i$ to TX $j$, and $l \in \{1,2,3,4\}, ~l \! \neq \! i$, denotes the index for the second RX unit on the same face as RX $i$. Fig. \ref{classicals} shows the configuration that these direct measurement algorithms are based on: One TX is in the field-of-view (FoV) of two RXs, and position is estimated over direct range or direct bearing estimates.

For bearing measurements, the positioning algorithm in \cite{soner_tvt} is utilized, which is based on triangulation with bearing measurements from two fixed anchors \cite{triangulation_bizimki}, i.e., it is the exact functional inverse of the system model equations that constitute its measurement set. The coordinate estimates, which are derived using the law of sines, are expressed as

\begin{equation}
	\label{p_j}
	\left[  
	\begin{matrix}
		\widehat{x_1}\\
		\widehat{y_1}
	\end{matrix}
	\right] = 
	\left[ 
	\begin{matrix}
		L\left(1 + \frac{\sin(\widehat{\theta_{21}})~\times~\cos(\widehat{\theta_{11}})}{\sin(\widehat{\theta_{11}} - \widehat{\theta_{21}})} \right)\\
		L\left(\frac{\cos(\widehat{\theta_{21}})~\times~\cos(\widehat{\theta_{11}})}{\sin(\widehat{\theta_{11}} - \widehat{\theta_{21}})} \right)
	\end{matrix}
	\right]~~,
\end{equation}

\noindent where $\widehat{\theta_{ij}}$ represents bearing measurements from RX $i$ to TX $j$, values $\widehat{x_1}, \widehat{y_1}$ are estimations for relative lateral and longitudinal position coordinates of TX 1 in the ego frame respectively, and $L$ is the known inter-RX distance on the ego vehicle. On the other hand for range measurements, the positioning algorithm in \cite{becha_positioning} is used, which is based on trilateration with the same setup as the bearing-based one in \cite{soner_tvt}, expressed as

\begin{equation}
	\label{p_j}
	\left[  
	\begin{matrix}
		\widehat{x_1}\\
		\widehat{y_1}
	\end{matrix}
	\right] = 
	\left[ 
	\begin{matrix}
		\frac{\left(\widehat{d_{11}}^2 - \widehat{d_{21}}^2 + L^2 \right)}{2L}\\
            \\
		\sqrt{\widehat{d_{11}}^2 - \widehat{x_1}^2}
	\end{matrix}
	\right]~~,
\end{equation}

\noindent where $\widehat{d_{ij}}$ represents range measurements from RX $i$ to TX $j$. Note that estimating position with respect to only one RX is sufficient since the two RX units on the ego vehicle are simply separated by $L$ and $L$ is known. Therefore $(x_{j},~y_{j}) \! = \! (x_{1j},~y_{1j})$ for TX $j$, $j \in \{1,2\}$.

\section{Newly Explored Algorithms}
We now formulate the two previously unexplored geometric algorithms for vehicular VLP: running position fixing using direct range measurements (i.e., single-receiver) and classical position fixing using differential bearing measurements. The new running fix algorithm is a range-based version of the running fix proposed in \cite{soner_pimrc} where only 1 RX unit is used for finding the position of a moving target vehicle when relative heading and speed information are known. Running fixes make it possible to locate the target vehicle even when it is at a very wide angle and one of the RXs lose the TX from its FoV (see \cite{running_fixing_1954} and \cite{soner_pimrc} for more info on running position fixing). The new classical position fixing using differential bearing measurements is a bearing-based version of the seminal work \cite{roberts_pdoa}, which was the state-of-the-art vehicular VLP method before the current state-of-the-art methods (\cite{soner_tvt} and \cite{becha_positioning}) were proposed. Hence, the newly explored algorithms are \quotes{siblings} of the existing algorithms \cite{soner_pimrc} and \cite{roberts_pdoa}, i.e., they have the same geometrical configuration, but bearing-based running position fix was explored in \cite{soner_pimrc} instead of range-based here, and vice versa for the other algorithm in \cite{roberts_pdoa}. Since \cite{soner_pimrc} and \cite{roberts_pdoa} have already been shown to be inferior to the current state-of-the-art methods recently, it is expected that the newly explored algorithms will perform similarly, i.e., inferior to the methods using direct measurements and classical position fixing (\cite{soner_tvt} and \cite{becha_positioning}). Nevertheless, we present these methods to motivate further research as well as to provide a complete picture of geometric algorithms in vehicular VLP literature.

\subsection{Running Fix, Direct Range Measurements} The geometry used for deriving the running fix using direct range measurements is given in Fig. \ref{new_mle}a. The algorithm first computes $\phi$ via the cosine theorem using range estimates $\widehat{d_v}, \widehat{d_{11}}(t_0)$ and $\widehat{d_{11}}(t_1)$. Subtracting $\left(180^\circ - \alpha_v \right)$ from $\phi$ gives $\beta$. Using $\beta$ and $\widehat{d_{11}}(t_0)$, the algorithm solves $\widehat{x_1}(t_0)$ and $\widehat{y_1}(t_0)$, i.e., the position at the first time instant. The algorithm then finds $(\widehat{x_1}(t_1),\widehat{y_1}(t_1)) - (\widehat{x_1}(t_0),\widehat{y_1}(t_0))$ using $\alpha_v$ and $d_v$, and using that relationship, solves for $\widehat{x_1}(t_1)$ and $\widehat{y_1}(t_1)$, i.e., the position at the second time instant. The full formulation of the algorithm is as follows:

\vspace{-2mm}
\begin{subequations}
	\label{mle_runrange}
	\begin{equation}
		\small
		\phi \!=\! \cos^{-1}\!\left( \frac{\! \widehat{d_v}^2 \!+\! \widehat{d_{11}}^2(t_0) \!-\! \widehat{d_{11}}^2(t_1) }{2\cdot \!\widehat{d_{v}} \!\cdot \!\widehat{d_{11}}(t_0) } \right)~,~ \beta \!=\! 180^\circ \!- \!\alpha_v
	\end{equation}	
	\vspace{0mm}
	\begin{equation}
		\small
		\left[  
		\begin{matrix}
			\widehat{x_1}(t_0)\\
			\widehat{y_1}(t_0) 
		\end{matrix}
		\right] \!=\! 
		\left[ 
		\begin{matrix}
			\widehat{d_{11}}(t_0) \! \cdot \! \cos \left(\beta\right) \\
			\widehat{d_{11}}(t_0) \! \cdot \! \sin \left(\beta\right)
		\end{matrix}
		\right]~~,
	\end{equation}
	\vspace{0mm}
	\begin{equation}
		\small
		\left[  
		\begin{matrix}
			\widehat{x_1}(t_1) \\
			\widehat{y_1}(t_1) 
		\end{matrix}
		\right] \!=\! 
		\left[ 
		\begin{matrix}
			\widehat{d_{v}} \! \cdot \! \cos \left(\alpha_v\right) \!+\! \widehat{x_1}(t_0) \\
			\widehat{d_{v}} \! \cdot \! \sin \left(\alpha_v\right) \!+\! \widehat{y_1}(t_0)
		\end{matrix}
		\right]~~.
	\end{equation}	
	\vspace{-2mm}
\end{subequations}

\noindent where $d_{v}$ and $\alpha_{v}$ are relative travelled distance and relative heading of the target vehicle with respect to the ego vehicle. These quantities are assumed to be computed based on the related speed and heading sensor measurements transmitted from the target to the ego vehicle.

\subsection{Classical Fix, Differential Bearing Measurements} 
\begin{figure}[t]
	\centering
	\includegraphics[width=0.50\textwidth]{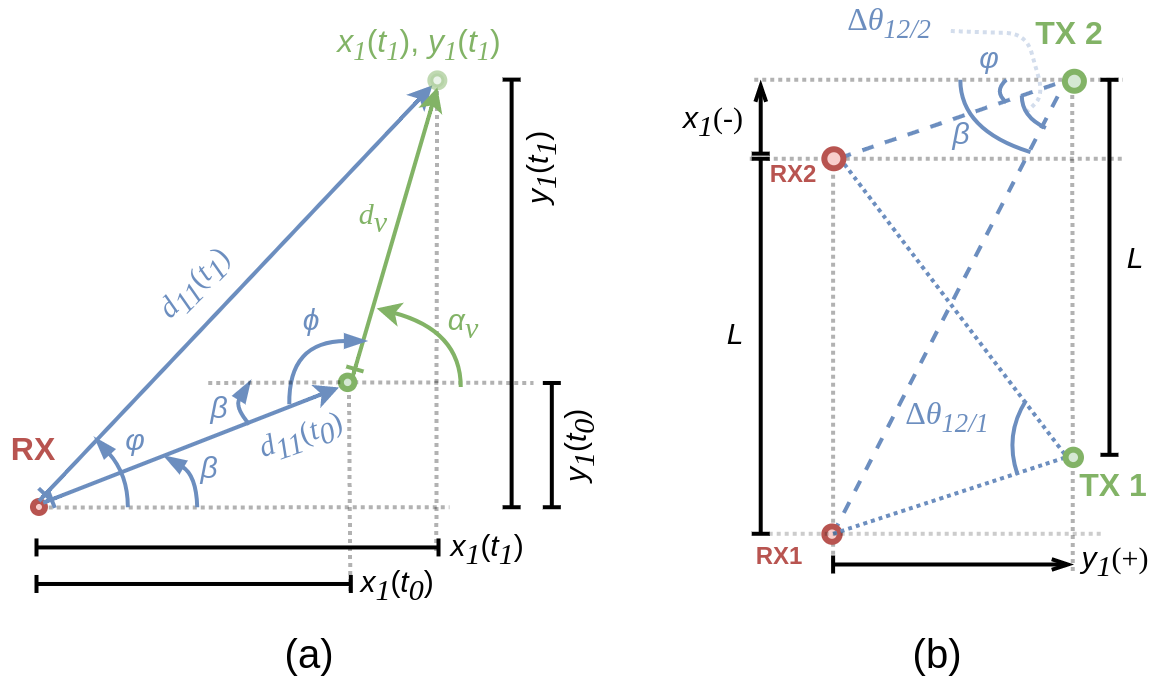}
	\vspace{-2mm}
	\caption{Geometries for (a) running position fixing using direct range measurements, and (b) classical position fixing using differential bearing measurements.}
	\vspace{-2mm}
	\label{new_mle}
\end{figure}

The vehicular VLP method in \cite{roberts_pdoa} solves the quadrilateral geometry formed by two RXs and two TXs that are parallel to each other and provides the highly complex position coordinate estimation expressions based on \quotes{delta lengths}, i.e., the differential range measurements, for classical position fixing using those measurements. In similar fashion, we solve the same quadrilateral geometry for angle-difference-of-arrival (ADoA, as in \cite[Fig. 1]{adoa_3d_vlp}), i.e., differential bearing measurements, and provide the associated positioning algorithm. The geometry is given in Fig. \ref{new_mle}b. The algorithm first defines \quotes{intermediate variables} $\varphi$ and $\beta$ shown in Fig. \ref{new_mle}b and expresses cotangents of angles $\Delta \theta_{12/1}$ and $\Delta \theta_{12/2}$ in terms of the unknown $x_1$ and $y_1$ using those two intermediate angles and the trigonometric identity for the addition of two inverse tangents. Then, assuming the RX1-RX2 and TX1-TX2 lines are parallel, $\sin(-\varphi)$ is derived from the two cotangents and afterwards used for computing $x_1$ and $y_1$. 	

The full formulation of the algorithm is as follows:

\begin{subequations}
	\vspace{1mm}
	\label{mle_difbear}
	\begin{equation}
		\varphi \!=\! \tan^{-1}\!\left(\frac{-x_1}{y_1}\right)~,~\beta \!=\! \tan^{-1}\!\left(\frac{L-x_1}{y_1}\right)~,
	\end{equation}
	\vspace{-1mm}
	\begin{equation}
		\Delta \theta_{12/2} = \beta - \varphi~,~\cot \left( \Delta \theta_{12/2} \right) \!=\! \frac{y_1^2 - L \cdot x_1 + x_1^2}{L \cdot y_1}~,
	\end{equation}
	\vspace{-1mm}
	\begin{equation}
		\cot \left( \Delta \theta_{12/1} \right) \!=\! \frac{y_1^2 + L \cdot x_1 + x_1^2}{L \cdot y_1}~,
	\end{equation}
	\vspace{-1mm}
	\begin{equation}
		\sin \left(-\varphi\right) = \frac{1}{2}\left(\cot \left(\Delta \theta_{12/2}\right) - \cot \left(\Delta \theta_{12/1}\right)\right)~,
	\end{equation}
	\vspace{-1mm}
	\begin{equation}
		\small
		\left[  
		\begin{matrix}
			\widehat{x_1}\\
			\widehat{y_1}
		\end{matrix}
		\right] = 
		\left[ 
		\begin{matrix}
			\sin \left(-\varphi\right) \cdot y_1 \\
			\left(\frac{L}{1+\left(\sin \left(-\varphi\right)\right)^2}\right)\left(\cot\left(\Delta \theta_{12/2}\right) - \sin \left(-\varphi\right)\right)
		\end{matrix}
		\right].
	\end{equation}
\end{subequations}
\vspace{1mm}

Considering this last formulation, it is important to emphasize that algorithms using differential measurements (\cite{roberts_pdoa} for range, and Eqn. \ref{mle_difbear} for bearing measurements) produce position estimates based on the assumption that the target and ego vehicles are parallel to each other. Therefore, even when this condition is not met, the methods interpret the input measurements as though they came from a target vehicle that is longitudinally parallel to the ego vehicle, and produce estimations accordingly. Although such an estimation obviously has systematic error when the vehicles are not parallel to each other, thus, results in biased estimates, this error can be acceptably small in practice for certain scenarios where heading difference between the two vehicles is small (e.g., during high speed cruise on a highway). For this reason, we do not discard the estimations from these methods during analyses even when the two vehicles are not longitudinally parallel, which happens very frequently in real driving scenarios due to high mobility. 

\section{Statistical Analyses}

In this section, we first show that geometric algorithms provide maximum likelihood estimates (MLE) for the respective sets of parameter measurements they utilize. Afterwards, we derive the Cramer-Rao lower bounds (CRLB) for all algorithms to determine the sensitivity of each algorithm against statistical VLC channel noise, for each position coordinate (x and y). Three classes of algorithms are considered: 1) classical fix using direct measurements, 2) classical fix using differential measurements and 3) running fix using direct measurements. Running fixes with differential measurements are not feasible for vehicular VLP since they would require the vehicle to move only sideways. Finally, we provide simulation results to verify the analyses and gather further performance insights.

\subsection{Maximum Likelihood Position Estimations}

Let $\mathbf{P}$ be a vector containing the true values of the quantities that are to be estimated (i.e., true position coordinates), $x_{j}~,~y_{j}~~j\in\{1,2\}$ for TX $j$. Let $\mathbf{G}$ be a vector containing the deterministic system model equations that relate the true values of quantities that are to be estimated to a set of measurements. Let $\mathbf{M}$ be a vector containing that set of $N_H$ measurements. Let $N_H$ be the number of measurements in that set. Let $\mathbf{W}$ be zero-mean additive white Gaussian noise (AWGN) terms with variance $\sigma_{\bf{W}}^2$ that contaminate the true values of those measurements, and let $M_h$, $G_h$ and $W_h$ be the element number $h$ for vectors $\mathbf{M}$, $\mathbf{G}$ and $\mathbf{W}$, respectively, where $h \in \{1,2,... , N_H\}$. This measurement set (i.e., set of $M_h$ values) can be expressed as 

\vspace{-1mm}
\begin{equation}
	\label{crlb_sys}
	M_h = G_{h}(\mathbf{P}) + W_h~~,~~h \in \{1,2,... , N_H\}~~.
\end{equation}
\vspace{-2mm}

\noindent Since $W_h$ are AWGN, $M_h$ are independent Gaussian random variables (r.v.) with mean $G_{h}(\mathbf{P})$ and variance $\sigma_{W_h}^2$. Thus, the log-likelihood function ($\ln p(\mathbf{M}; \mathbf{P})$), which is the logarithm of the joint probability density function (PDF) for the r.v.s, is

\vspace{-2mm}
\begin{equation}
	\label{likelihood}
	\small
	\ln p(\mathbf{M}; \mathbf{P}) = - \frac{\sum_{h=1}^{N_H} \ln 2 \pi \sigma_{W_h}^2}{2} - \sum_{h=1}^{N_H} \frac{ \left( M_h - G_{h}(\mathbf{P}) \right)^2}{2 \sigma_{W_h}^2}~~.
\end{equation}

\noindent MLE occurs at the global maximum of this unimodal distribution, which is at the position that makes the first partial derivatives of the log-likelihood function with respect to each quantity equal to 0. Thus, the MLE positioning algorithm for a given measurement set can be formulated by solving the following for each element in vector $\mathbf{P}$:

\vspace{-2mm}
\begin{equation}
	\label{likelihood_turev}
	\frac{\delta \ln p(\mathbf{M}; \mathbf{P})}{\delta P_m}  = \sum_{h=1}^{N_H} \frac{M_h - G_{h}(\mathbf{P})}{\sigma_{W_h}^2} \cdot \frac{\delta G_{h}(\mathbf{P})}{\delta P_m} = 0~,
\end{equation}

\noindent where $P_m, ~m \in \{1,2,...,N_M\}$ is $m^{th}$ element in vector $\mathbf{P}$, i.e., the true value for the $m^{th}$ quantity that is to be estimated, and $N_M$ is the number of quantities to be estimated. Conversely, an algorithm can be shown to provide MLE by simply replacing $P_m$ with $\widehat{P_m}$, the estimation for the $m^{th}$ quantity, and showing that Eqn. (\ref{likelihood_turev}) holds. We next demonstrate this for the state-of-the-art vehicular VLP method for direct bearing based classical position fixing \cite{soner_tvt} as an example. 

The positioning algorithm in \cite{soner_tvt} is based on triangulation with bearing measurements from two fixed anchors \cite{triangulation_bizimki}, i.e., it is the exact functional inverse of the system model equations that constitute its measurement set. Note that this is possible due to $N_H = N_M$, i.e., due to the set of equations being determined. The system model equations are the geometric relations that tie the position coordinates to the range and bearing measurements, expressed as follows

\vspace{-1mm}
\begin{equation}
	\label{rx_model_phs}
	\small
	d_{ij} \!=\! \sqrt{{x_{ij}}^2 \! + \! {y_{ij}}^2}~~,~~\theta_{ij} = \arctan \left( \frac{x_{ij}}{y_{ij}} \right)~~.
\end{equation}
\vspace{-1mm}

\noindent where $\theta_{ij}, d_{ij} ~~i,j\in\{1,2\}$ represent bearing and range from RX $i$ to TX $j$ respectively.

Using Eqn. (\ref{rx_model_phs}) to form the $G_{h}(\mathbf{P})$ terms, replacing the $(x_1, y_1)$ terms in those equations with $(\widehat{x_1}, \widehat{y_1})$ from Eqn. (\ref{p_j}), and noting that $M_h$ terms correspond to $\widehat{\theta_{11}}$ and $\widehat{\theta_{21}}$, the first term of the multiplication inside the sum in Eqn. (\ref{likelihood_turev}) becomes 0 for all $h \in \{1,2\}$ after simplifications, thus, proves that the algorithm in Eqn. (\ref{p_j}) provides MLE for its measurement set since Eqn. (\ref{likelihood_turev}) holds. Following from this demonstration, we make the following generalization here, which holds for both the other state-of-the-art vehicular VLP geometric algorithm (direct range-based classical fix) as well as the newly proposed algorithms: For a set of measurements that provides a determined set of equations (i.e., $N_H = N_M$), if the position estimation algorithm is formulated using the functional inverse of the related system model equations, that algorithm provides maximum likelihood position estimations as long as the measurement set can be expressed with Eqn. (\ref{crlb_sys}), i.e., when parameter measurements are contaminated with AWGN. Since previous works in the literature show this to be true, the generalization holds \cite{soner_pimrc, soner_tvt, kay, smartAutoLighting, becha_vlr_experimental}.

\subsection{Cramer-Rao Lower Bounds}

We now derive the CRLBs on positioning accuracy for the state-of-the-art positioning algorithms. The CRLB lower bounds the mean-squared-error (MSE) for an unbiased estimate of a quantity considering related noisy measurements. It is a suitable tool for analytically evaluating the performance of a vehicular VLP positioning algorithm under noise since individual parameter measurements in vehicular VLP are contaminated by AWGN due to channel noise of the same statistical nature \cite{soner_pimrc, soner_tvt, kay, smartAutoLighting, becha_vlr_experimental}. For this reason, we derive the bound for each algorithm here with respect to their bearing/range measurements. The resulting CRLB thus expresses the robustness of unbiased estimates against errors in parameter measurement, and is applicable to all individual parameter measurement techniques, including those proposed for different RX designs. 

The derivation procedure \cite{soner_tvt} first computes the Fisher information matrix (FIM), $\bf{F}$, for the considered measurement set (a.k.a. the \quotes{dataset} \cite{kay}). The diagonal elements of the inverse of the FIM correspond to the lower bound of MSE (i.e., minimum variance around the true value) for each estimation:

\vspace{0mm}
\begin{equation}
	\label{crlb_eqn}
	var(\widehat{P_m}) \geq {\left({\mathbf{F}^{-1}}\right)}_{m,m'}~,
\end{equation}
\vspace{-2mm} 

\noindent where $var()$ denotes variance, $m, m' \in \{1,2,..., N_M\}$ denote row and column indices for the FIM, respectively, and $m=m'$ in Eqn. (\ref{crlb_eqn}) (recall that $N_M$ is the number of quantities that are to be estimated and that $\widehat{P_{m}}$ is the estimation for the $m^{th}$ quantity). The elements of the FIM are computed by:

\vspace{1mm}
\begin{equation}
	\label{crlb_fim_elts}
	\small
	\mathbf{F}_{m,m'} = -\sum\limits_{h=1}^{N_H} \frac{1}{\sigma_{W_h}^2} \left( \frac{\delta G_{h}(\mathbf{P})}{\delta P_{m}} \cdot ¨\frac{\delta G_{h}(\mathbf{P})}{\delta P_{m'}} \right)~,
\end{equation}
\vspace{1mm}

\noindent where $\sigma_{W_h}^2$, $ G_{h}$ and $\mathbf{P}$ were defined in Section III-A. Since the final symbolic CRLB expressions (i.e., after inverting the FIM in Eqn. \ref{crlb_eqn}) are too complex to the degree that they do not provide any additional intuition, we present only the computation of the FIM elements during the derivations based on Eqn. \ref{crlb_fim_elts}, and the final CRLBs are numerically computed in Section IV during simulations using the terms provided here. For each algorithm, we first point to the relevant system model equations that constitute the $G_h()$, $h \in \{1,2,..., N_H\}$ functions, and then provide the $\delta G_h() / \delta x_{1}$ and $\delta G_h() / \delta y_{1}$ terms for each of the $N_H$ measurements, which are used for building the FIM via Eqn. (\ref{crlb_fim_elts}).


\vspace{2mm}
\subsubsection{CRLB, Classical Fix, Direct Measurements} 
These algorithms use measurements ($\widehat{d_{11}}$, $\widehat{d_{21}}$) for range and \linebreak($\widehat{\theta_{11}}$, $\widehat{\theta_{21}}$) for bearing. These are measurements for the actual direct range and bearing values from TX 1 to RX 1 and to RX 2 respectively (cf., Eqn. (\ref{rx_model_phs})), contaminated by AWGN of variance ($\sigma_{d_{11}}^2$, $\sigma_{d_{21}}^2$) for range and ($\sigma_{\theta_{11}}^2$, $\sigma_{\theta_{21}}^2$) for bearing, respectively. Noting that $N_H \!=\! N_M \!=\! 2$, i.e., that a minimal configuration is present (2 RXs and 2-3 TXs on each face), the derivative terms necessary for constructing the associated FIM using Eqn. (\ref{crlb_fim_elts}) for range measurements are:

\vspace{1mm}
\begin{subequations}
	\begin{equation}
		\label{classical_direct_range1}
		\frac{\delta d_{11}}{\delta x_{1}}= \frac{x_1}{\sqrt{x_{1}^2 + y_{1}^2}}~,~\frac{\delta d_{21}}{\delta x_{1}}= \frac{(x_1 - L)}{\sqrt{(x_{1}-L)^2 + y_{1}^2}}
	\end{equation}
	\vspace{-1mm}
	\begin{equation}
		\label{classical_direct_range2}
		\frac{\delta d_{11}}{\delta y_{1}}= \frac{y_1}{\sqrt{x_{1}^2 + y_{1}^2}}~,~\frac{\delta d_{21}}{\delta y_{1}}= \frac{y_1}{\sqrt{(x_{1}-L)^2 + y_{1}^2}}.
	\end{equation}
\end{subequations}
\vspace{1mm}

\noindent Repeating this for the bearing measurements yields:

\vspace{1mm}
\begin{subequations}
	\begin{equation}
		\label{classical_direct_bearing1}
		\frac{\delta \theta_{11}}{\delta x_{1}}= \frac{y_1}{x_{1}^2 + y_{1}^2}~,~\frac{\delta \theta_{21}}{\delta x_{1}}= \frac{y_1}{(x_{1}-L)^2 + y_{1}^2}
	\end{equation}
	\vspace{1mm}
	\begin{equation}
		\label{classical_direct_bearing2}
		\frac{\delta \theta_{11}}{\delta y_{1}}= -\frac{x_1}{x_{1}^2 + y_{1}^2}~,~\frac{\delta \theta_{21}}{\delta y_{1}}= -\frac{(x_1-L)}{(x_{1}-L)^2 + y_{1}^2}.
	\end{equation}
\end{subequations}

\vspace{4mm}
\subsubsection{CRLB, Classical Fix, Differential Measurements}
These algorithms use measurements ($\widehat{\Delta d_{12/1}}$, $\widehat{\Delta d_{12/2}}$) for range and ($\widehat{\Delta \theta_{12/1}}$, $\widehat{\Delta \theta_{12/2}}$) for bearing. These are measurements for the differences in actual range/bearing values of TX 1 to \linebreak RX 1 and RX 2 and the same for TX 2 as governed by Eqn. (\ref{diffs}). These measurements are contaminated by AWGN of variance ($\sigma_{\Delta d_{12/1}}^2$, $\sigma_{\Delta d_{12/2}}^2$) for range and ($\sigma_{\Delta \theta_{12/1}}^2$, $\sigma_{\Delta \theta_{12/1}}^2$) for bearing, respectively. Note that $N_H \!=\! N_M \!=\! 2$, and that the model only allows unbiased estimates when \linebreak $(x_{2}, y_{2}) = (x_{1} + L, ~y_{1})$, i.e., when the target vehicle and ego vehicle are longitudinally parallel to each other. 

The derivative terms necessary for constructing the associated FIM using Eqn. (\ref{crlb_fim_elts}) for range measurements can be computed using exactly the same procedure used for that of the direct measurements.

\vspace{2mm}
\subsubsection{CRLB, Running Fix, Direct Measurements}
The running position fix is fundamentally different from the classical fix: It explicitly requires target vehicle relative movement since it is formulated for estimation of position at two  consecutive time steps (4 quantities in total), i.e., for $\mathbf{P} = \left[x_{1}(t_0)~~y_{1}(t_0)~~x_{1}(t_1)~~y_{1}(t_1)\right]$, where $t_0$ and $t_1$ denote the two consecutive time instants separated by the VLP update period. Thus, it requires two extra measurements for a determined system: The relative target vehicle heading ($\alpha_v$) and distance traveled ($d_v$). The relations between $\alpha_v$ and $d_v$ and the position are as follows:

\begin{subequations}
	\begin{equation}
		\label{alpha_v_eq}
		\alpha_v = \arctan \left( \frac{x_{1}(t_1) - x_{1}(t_0)}{y_{1}(t_1) - y_{1}(t_0)} \right)
	\end{equation}
	\vspace{1mm}
	\begin{equation}
		\label{d_v_eq}
		d_v = \sqrt{(x_{1}(t_1) - x_{1}(t_0))^2 + (y_{1}(t_1) - y_{1}(t_0))^2}
	\end{equation}
	\vspace{-1mm}
\end{subequations}

\noindent where, ideally, $a_v$ and $d_v$ should be constant between time instants $t_0$ and $t_1$ \cite{soner_pimrc}. The measurements for $\alpha_v$ and $d_v$, i.e., $\widehat{\alpha_v}$ and $\widehat{d_v}$ respectively, can be computed from on-board inertial sensor information transmitted over the VLC channel from the target to the ego. The measurements can be assumed to be contaminated by AWGN and standard deviation values can be obtained from sensor datasheet information \cite{soner_pimrc}. These are combined with two consecutive direct measurements of bearing ($\widehat{\theta_{11}}(t_0)$ and $\widehat{\theta_{11}}(t_1)$) or range ($\widehat{d_{11}}(t_0)$ and $\widehat{d_{11}}(t_1)$) for position estimation (i.e., $N_H \!=\! N_M \!=\!4$ in both cases). The derivative terms for the direct bearing and range measurements were already derived in Eqns. (\ref{classical_direct_bearing1}) and (\ref{classical_direct_bearing2}) and Eqns. (\ref{classical_direct_range1}) and (\ref{classical_direct_range2}) respectively, and the derivation of the terms for $\alpha_v$ and $d_v$ are straightforward using the same primitives.

Before concluding this section, we also discuss two more statistical properties that emerge from the CRLBs, common for all of the MLE positioning algorithms presented above: 1) At extremely low signal-to-noise-ratio (SNR), the symmetrical Gaussian distributions at the algorithm inputs (i.e., parameter measurements) naturally get transformed into asymmetrical ones at the outputs (i.e., position estimations) due to the non-linearity of the algorithms, making the position estimations biased. However, as discussed in detail in \cite[Ch. 7.2]{kay}, estimates are still strongly unbiased at moderate to high SNR, which reinforces the usefulness of the CRLB analyses under most conditions. Moreover, 2) these estimators converge to the minimum variance unbiased estimators (MVUE) for their respective problems at moderate to high SNR. While showing this via the CRLB theorem, i.e., by showing that the algorithm expressions satisfy \cite[Eqn. (3.7)]{kay}, is not straightforward due to the non-linearity involved, it is straightforward to show this using the Rao-Blackwell-Lehmann-Scheffé (RBLS) theorem \cite{kay}. \linebreak The proof for each algorithm is the same, and follows from \cite[Example 5.6]{kay} by noting the following: 

\vspace{1mm}
\begin{itemize}
	\item the estimator is unbiased for practical SNR levels,
	\vspace{1mm}
	\item the input parameter measurement set (i.e., \quotes{dataset}) itself is a sufficient statistic for the problem, and
	\vspace{1mm}
	\item this sufficient statistic is also complete \footnote{The set of input parameter measurements for the algorithms is not exactly of the same nature as the \quotes{dataset} in \cite[Example 5.6]{kay}: The algorithms here are all nonlinear and in vector form. However, Gaussian parameter noise PDFs ensure that those statistics are also complete, like in \cite[Example 5.6]{kay}}.
\end{itemize}
\vspace{1mm}

\noindent The preceding holds for all algorithms, thereby proving that they converge to the MVUEs for their respective problems for the considered practical SNR regime. We next evaluate these deductions about statistical performance via simulations.

\subsection{Simulations}

The simulation setup considers typical passenger cars with different dimensions as ego and target vehicles. The target vehicle is considered to be transmitting from its tail light towards the ego vehicle like shown in Fig. \ref{classicals}b to evaluate the worst case scenario (taillights have lower power than headlights). The taillight has 2 W optical power, and a Lambertian pattern with 20\textdegree ~half-power angle (order m=11) to match the TX configuration used in \cite{becha_positioning}. TX signal is a 1 MHz pure tone like in \cite{becha_ranging}, which is band-pass filtered at 100 kHz bandwidth around the 1 MHz carrier on the RX end as in \cite{becha_vlcr_experimental}. 100 Hz positioning rate is considered, analog-to-digital conversion occurs at 10 MSPS, and all methods utilize the \quotes{QRX} unit \cite{soner_tvt}: range-based methods use the sum of the RX signals. The QRX uses an Edmund Optics \#67-149 plano-convex lens, providing $\approx\pm60^\circ$ FoV. Clear weather and indirect sunlight exposure is assumed on the QRX. Further details on the simulation setup are presented on the GitHub repository containing the complete Python source code \cite{github_vvlp}.

We first simulate the performance of both the two newly explored methods as well as the current state-of-the-art methods in a dynamic lane change scenario. Afterwards, we characterize the theoretical performance of the two state-of-the-art methods over their feasible operational range by evaluating their CRLBs over a grid that covers a 3-lane road to assess their sensitivity against statistical channel noise. 

\begin{figure}[t]
	\centering
	\includegraphics[width=0.50\textwidth]{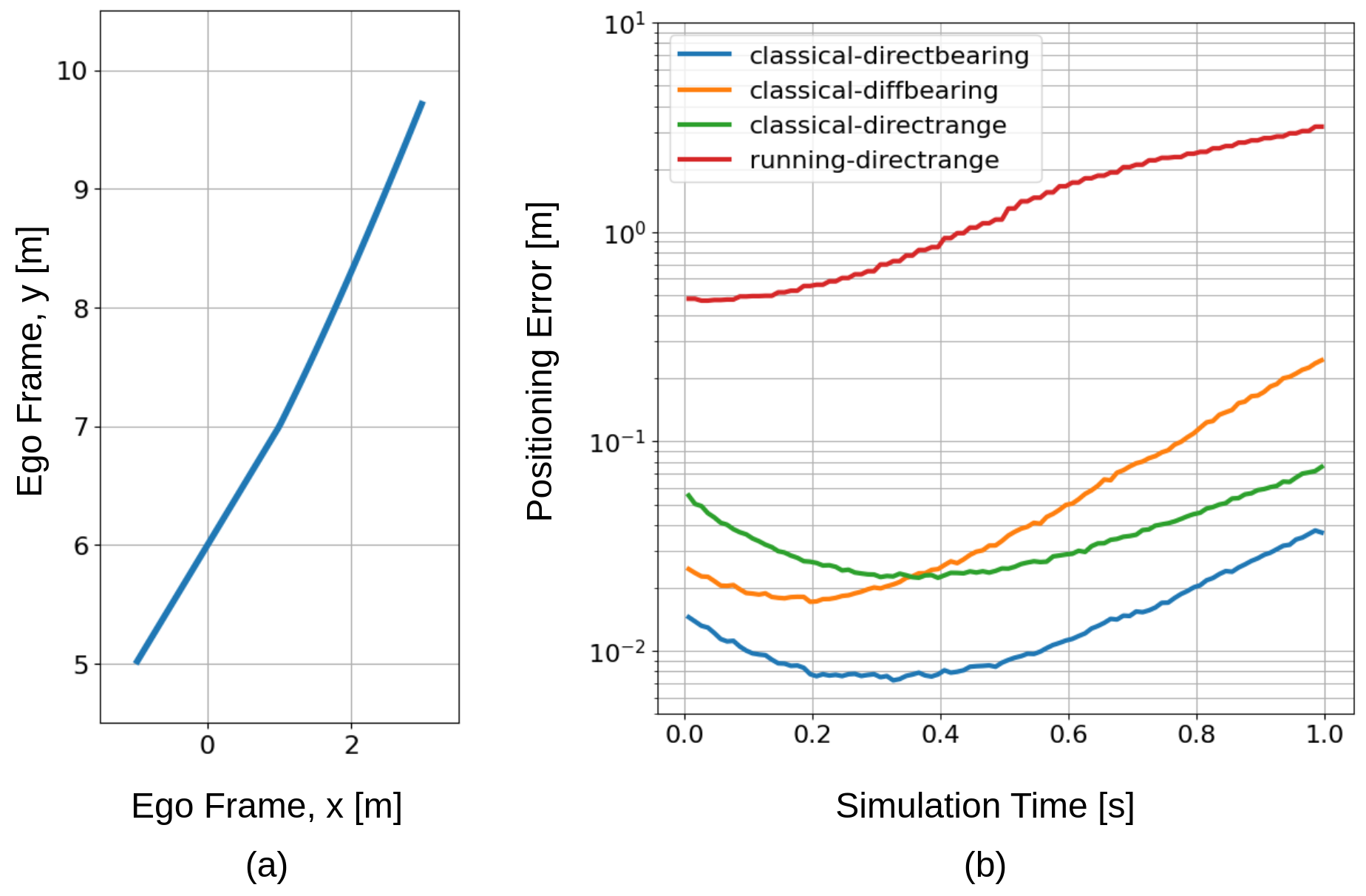}
	\vspace{-2mm}
	\caption{(a) Trajectory of the lane change simulation, and (b) positioning error for methods over the trajectory.}
	\vspace{-2mm}
	\label{report_sim1}
\end{figure}

Fig. \ref{report_sim1}a depicts the relative trajectory of the target vehicle within the ego frame of reference for the lane change trajectory and Fig. \ref{report_sim1}b shows the standard deviation of the 2D positioning error each algorithm makes over this trajectory, aggregated over 3000 iterations of the same simulation with different random noise samples. The results verify our initial hypothesis: the two newly explored methods, namely, differential bearing-based classical fix and direct range-based running fix, are inferior to the current state-of-the-art, i.e., direct measurement-based classical position fixing methods. 

\begin{figure}[t]
	\centering
	\includegraphics[width=0.50\textwidth]{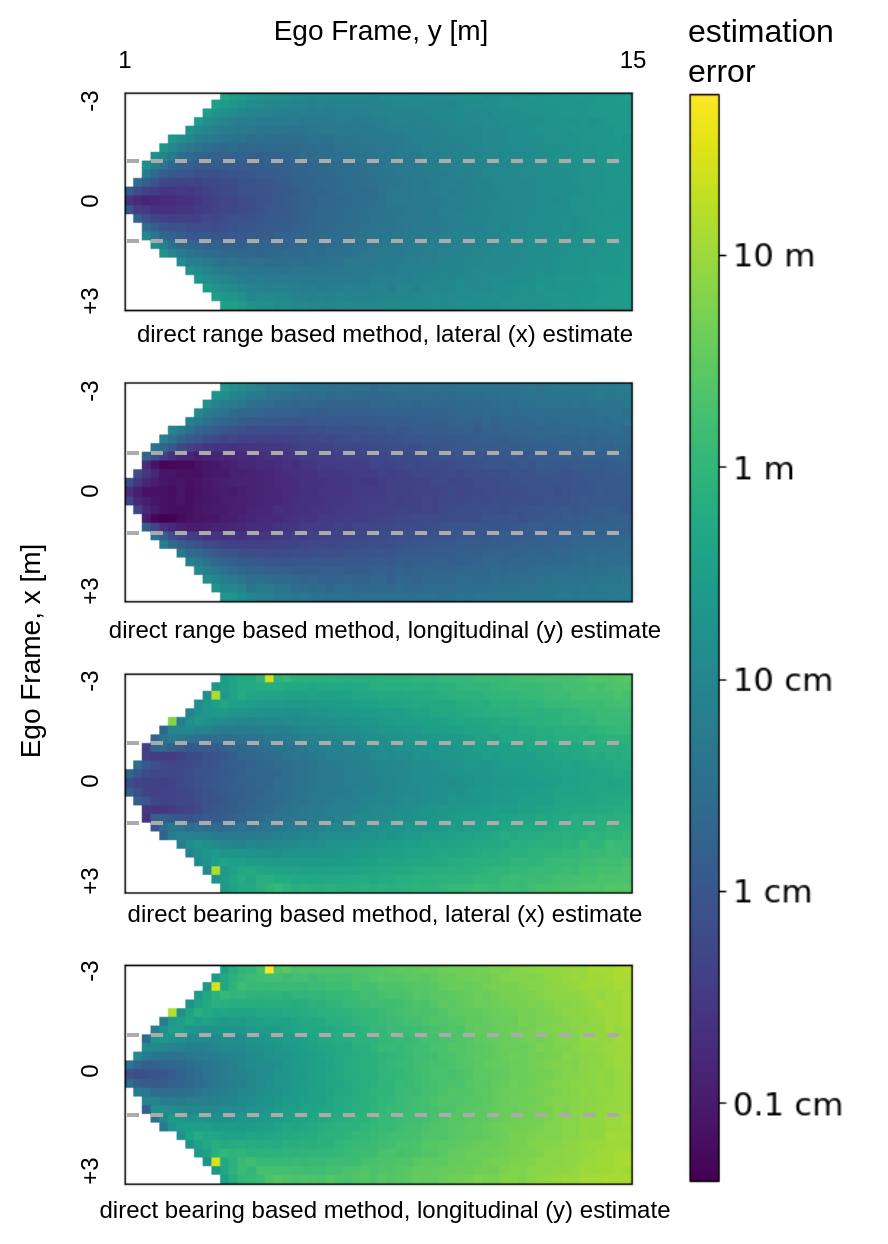}
	\vspace{2mm}
	\caption{Sensitivity analyses for the methods utilizing direct bearing and range measurements on the lateral (x) and longitudinal (y) axes.}
	\vspace{2mm}
	\label{report_sim2}
\end{figure}

The two direct measurement-based state-of-the-art methods are also observed to perform very differently in Fig. \ref{report_sim1}b with respect to the parameter that is being measured, i.e., bearing or range. To characterize this difference, we evaluate the theoretical performance of both methods over their complete operational range. To this end, we sample all positions over a grid that covers the distance of a 3-lane highway road, simulate signal propagation, sample errors in bearing and range parameter measurements over many iterations to get aggregate statistics against channel noise (\cite{soner_tvt} for bearing and \cite{roberts_pdoa} for range), and obtain bearing and range measurement error distributions for each point. The standard deviation of each distribution corresponds to quantity $\sigma_{W_h}$ in Eqn. \ref{crlb_fim_elts}. We evaluate the CRLB for each location, using each $\sigma_{W_h}$ measurement to obtain the lower bound on positioning error for that location, we repeat this over the whole grid to get an \quotes{error map} over the 3-lane road. The results for this simulation, shown in Fig. \ref{report_sim2}, demonstrates a clear difference between the sensitivities of bearing and range based classical position fixing methods against statistical channel noise. Specifically, bearing-based classical position fixing provides higher accuracy for estimation in the lateral axis (x in ego frame of reference) compared to the longitudinal axis (y) and range-based classical position fixing provides higher accuracy for estimation in the longitudinal axis (y). This result motivates investigating hybrid methods that utilize both bearing and range measurements if receiver architectures that can realize both methods simultaneously get realized. 

\section{Conclusion}

In this paper, we explored two geometric algorithms that were previously unexplored in vehicular VLP, analyzed all geometric algorithms in the literature by deriving CRLBs and investigating whether or not these algorithms provide MLE and MVUE to characterize their theoretical performance, and simulated the performance of both the newly explored algorithms as well as the state-of-the-art methods. Results show that the two newly explored algorithms are inferior to the current state-of-the-art methods which utilize classical position fixing using direct bearing-range measurements. Our main finding was that the sensitivity of the state-of-the-art methods for positioning in the lateral and longitudinal axes differed significantly based on which parameter was being used for measurement. Specifically, direct bearing-based positioning provides higher accuracy in the lateral axis, and direct range-based positioning provides higher accuracy in the longitudinal axis. The results motivate future work on vehicular VLP using combination of bearing and range parameter measurements with receiver architectures that can realize both measurements simultaneously.



\bibliographystyle{IEEEtran}
\bibliography{sensitivityreport}

\end{document}